# Impacts of real-time data collection on introductory algebra-based physics

Eric Brewe, Vashti Sawtelle, and Priscilla Pamela

Florida International University

Including real-time data collection technology is a common practice to upgrade physics labs, and the assumption is such technology improves student learning, yet little has been done to demonstrate the effects of technology.  Specific activities have been shown to be enhanced by technology, but the effects on the class as a whole has been left unexplored. This paper investigates the effects of technology on two algebra based introductory physics classes. In this paper, we use FCI, MPEX, surveys, and ethnographies to document the improvement in learning as real-time data collection technology is incorporated into a modeling physics class. The two classes examined differed only in the inclusion of technology. The results found were significant. Students in the class with high technology were found to have learned more than students in the class with no technology. This paper explores the gains in learning and relates them directly to the addition of technology.

## Introduction

The impacts of technology on education are undeniable. Nowhere is this as evident as in the physics classroom. Technological innovations allow physics teachers many instructional options including real-time data collection, numerical analyses of realistic problems, and simulation of complex phenomena. Inclusion of these instructional tools implies that technology enhances student learning.[1] Research has attributed learning gains to technology on specific concepts within the physics curriculum.[2,3,4]  The impacts of technology on learning in an entire course are harder to isolate and less well defined. This article looks at both evidence of the impacts of technology on learning in an introductory algebra-based mechanics course and then learning in the specific context of momentum.

This research is supported by a Course, Curriculum, Lab and Instruction grant from the National Science Foundation, "Re-Modeling for Algebra-based Physics."[a] The primary purpose of the CCLI program is to support adaptation and implementation of educational reforms.[5] The Re-modeling Algebra-based Physics (RAP) project is built on the foundation of the Re-modeling University Physics project[b] from Arizona State University. The RAP project took place at Hawaii Pacific University, a medium-sized liberal arts university in Hawaii. The Re-modeling project at ASU adapted the Modeling approach to physics instruction to calculus-based physics and the Re-modeling project at HPU focused on adapting to algebra-based physics.

Hands-on, studio-format classes, with the "lab" and "lecture" components integrated, characterize Modeling instruction.. The instruction is student-centered and focuses creating situations where students learn to construct and validate models.[6,7,8] A key element in the construction and validation of models is the correspondence of the model to actual physical phenomenon. Ideally, high-tech data acquisition hardware and software supports the development of models through real time data collection and representation. Student activity centers on interpreting and analyzing the data they have collected.

The primary goal of the Re-modeling Algebra-based Physics project was to support the adaptation of labs from no-tech to high-tech. Thus, the only significant difference between the two classes being studied here, which took place during Fall 2004 and Fall 2005, is the inclusion of high tech labs supported by Pasco™ data acquisition hardware and data analysis software. The course format, pedagogy, instructor, student population, even class meeting time remained the same. The labs were all modified from existing no-tech hands-on labs focused on developing conceptual understanding to high-tech hands-on labs focused on developing conceptual understanding. As a result, the design of the experiment allows for measurement of the impact of technology on learning in this context.

**Experimental Design**

This study took place in an intact algebra-based introductory mechanics course (PHYS 2030) during two subsequent years, Fall 2004 (no technology) and Fall 2005 (with technology). Like other classes at liberal-arts universities, the physics classes are small, averaging around 25 students, with a maximum of 30. The algebra-based physics class serves biology, computer science, environmental studies and pre-medical majors and like many classes at HPU is made up of students from a variety of ethnicities. The first author on this paper was the instructor for these courses.

---

**Table I. Demographics of students enrolled in Phys 2030 during Fall 2004 and 2005**

|  |  | Fall 2004 |  | Fall 2005 |
|---|---|---|---|---|
| Gender | Female | 56.8% | Female | 57.9% |
|  | Male | 43.1% | Male | 42.1% |
| Ethnicity | White | 47.0% | White | 52.6% |
|  | Asian | 7.8% | Asian | 18.4% |
|  | Pacific Islander | 33.3% | Pacific Islander | 10.5% |
|  | Black | 3.9% | Black | 5.2% |
|  | Native American | 0% | Native American | 2.6% |
|  | Hispanic | 5.8% | Hispanic | 2.6% |
|  | Other | 1.9% | Other | 7.9% |

In order to establish the differences in learning between the two subsequent years and to document factors leading to these differences, the first author collected three forms of data.

*FCI and MPEX*

First, quantitative pre/post data from the Force Concept Inventory (FCI) and Maryland Physics Expectations survey (MPEX) was collected to assess conceptual learning and to determine attitudinal shifts in students.[9] The normalized gain <g>, or Hake gain was calculated and used as a comparison statistic on the FCI scores.[10]

*Student Surveys*

Student surveys, administered through WebCT post instruction were used to gauge students' reactions to the inclusion of technology and to characterize their use and familiarity with the technology. The surveys, could not be identical, because it would be impossible for students without technology to gauge how technology would have helped, but it is plausible students with technology could gauge how the technology helped in comparison with other science courses that did not include technology. As a result, the student surveys were analyzed by two Physics Education Researchers (PEResearchers). Student responses were characterized as either (F) favorable toward use of technology, (U) unfavorable toward the use of technology or (N) neutral toward the use of technology.

*Ethnographies*

The third data source was ethnographies written by a supporting faculty member visiting the class for the momentum labs in each of the two years. Ethnography is a valuable tool for providing an external characterization of the differences in instruction in the two years. The first ethnography was completed in Fall 2004 and to preserve objectivity, the first author did not read the ethnography until after the second ethnography had been completed in Fall 2005. The ethnographies were read by the authors who identified the elements of the ethnography that related to student use of technology. Then the elements relating to technology were compared between the two years and quotes were selected to demonstrate differences.

## Results and Analysis

The two classes, 2004 (no tech) and 2005 (tech) were compared on pretest FCI scores using independent samples t-test, $t(87) = 1.23$, $p = .220$, indicating that the two classes started with equivalent understanding of Newtonian Mechanics. However, the two classes differed significantly on Posttest score $t(87) = 2.80$, $p < .01$ and Hake gain $t(87) = 2.90$, $p < .01$. Because the two classes differ significantly on FCI post, and <g> it can be inferred that the 2005, technology enhanced class, learned significantly more than the 2004, no-technology class. The effect size, d, was calculated for each of the two t-tests; the effect size for the FCI post, d = 2.13 and the effect size for the <g>, d = 4.06 both indicate large effects, using the convention proposed by Cohen, d > 0.8 are considered large effects.[11] The interpretation of these statistics is clear, students learned more in a technology enhance class than in a class without technology.

**Table II. FCI Scores and Hake gain**

|  | Fall 2004 | Fall 2005 |
|---|---|---|
| FCI Pretest | 7.12±3.39 | 8.08±3.98 |
| FCI Posttest | 12.07±5.34 | 15.93±7.43 |
| Gain <g> | 0.224±0.195 | 0.383±0.305 |

The question turns then to why did students learn more in the technology enhanced class? Together, the results from the ethnographies and the MPEX, may indicate that there were significant differences in the pedagogy between the two classes.

### MPEX Results

The MPEX was used to determine the role of the pedagogy on the students. For the purposes of this study if the overall pedagogy did not change, we expect to see no differences between the MPEX profiles from year to year. As can be seen in Table III, the modeling course produces increases in all MPEX categories except effort, and the increases are similar in both years of the study, which indicates that the structure of the classes and the pedagogy employed were consistent throughout.

**Table III. MPEX pre and post scores, Fall 2004, Fall 2005.**

|  | 2004 Pre | 2004 Post | 2005 Pre | 2005 Post |
|---|---|---|---|---|

| | | | | | | | | |
|---|---|---|---|---|---|---|---|---|
| Overall | Fav. | 55.2 | Fav. | 62.8 | Fav. | 56.8 | Fav. | 62.0 |
| | Un. | 24.4 | Un. | 21.2 | Un. | 23.0 | Un. | 20.6 |
| Independence | Fav. | 44.1 | Fav. | 57.4 | Fav. | 53.2 | Fav. | 54.4 |
| | Un. | 38.1 | Un. | 31.6 | Un. | 29.4 | Un. | 30.0 |
| Coherence | Fav. | 46.2 | Fav. | 53.1 | Fav. | 51.1 | Fav. | 54.2 |
| | Un. | 35.6 | Un. | 29.4 | Un. | 28.1 | Un. | 26.2 |
| Conceptual | Fav. | 50.7 | Fav. | 66.4 | Fav. | 52.8 | Fav. | 64.9 |
| | Un. | 25.8 | Un. | 19.1 | Un. | 30.6 | Un. | 17.8 |
| Reality | Fav. | 68.9 | Fav. | 75.5 | Fav. | 65.9 | Fav. | 73.9 |
| | Un. | 11.1 | Un. | 5.9 | Un. | 9.6 | Un. | 11.1 |
| Math | Fav. | 57.3 | Fav. | 66.4 | Fav. | 60.9 | Fav. | 66.7 |
| | Un. | 20.4 | Un. | 15.7 | Un. | 20.9 | Un. | 16.0 |
| Effort | Fav. | 74.5 | Fav. | 64.7 | Fav. | 73.6 | Fav. | 68.0 |
| | Un. | 11.6 | Un. | 19.6 | Un. | 14.0 | Un. | 18.7 |

**Ethnography Results**

The ethnographer that visited the class in each of the two years produced field notes that can be used to look for examples of the differences in the teaching and to identify reasons for the increased learning with technology. The lab was on collisions and momentum transfer. During Fall 2004, the lab used metal ball bearings rolling down tracks and colliding. In keeping with the modeling approach, students had created a model to explain the situation and used the model to make predictions before the collisions, then they collided the marble, and finally they used their findings to revise their model. Due to the lack of equipment students made visual observations and then recorded the observations using motion maps for the two carts as data. The lab was very similar in Fall 2005. Students made models on whiteboards, which they then used to generate predictions about the outcome of the collisions. The primary difference during Fall 2005 was the inclusion of the technology, instead of colliding ball bearings, students used Pasco PasCarts and collected data using motion detectors. Laptop computers then instantly displayed position vs. time and velocity vs. time graphs. Fall 2004
The ethnographer made comments that fell into two categories during Fall 2004. The first group of comments related to the difficulty of using tracks and ball bearings, citing problems such as, "the track was wobbly, and the balls don't stay together." The second group of comments related to making observations as a means for collecting data. One student reported she was, "observing the force of the collisions," while another pointed out that they had made their predictions, but after trying to use the equipment, they had to simplify their observations.

*Fall 2005 Ethnography*

When the ethnographer returned in Fall 2005, the technology related comments were again in two categories, but they were different than in Fall 2004. In the first category were comments relating to difficulties with the technology such as, "[there was]…some initial difficulty getting the laptops set up for the experiment," and "the students would not be able to retrieve the data outside of class". The second group of comments related to how the students were interacting with the data. She explained how one group was working on understanding the results in terms of velocity of the two carts (one is negative and one is positive), and in another case that students described, "using position over time to compute velocity, which was graphed on the laptop for each trial," they were able to generate predictions about the outcomes.

Notable in these two groups of comments, is that for Fall 2005 the comments were not all positive about technology, there are barriers to overcome, many of which are unforeseen. However, the students were all able to overcome the barriers, and were able to complete the lab activity. Also the ethnographer indicated in her notes that the lab lasted only an hour, whereas during Fall 2004, the lab lasted a full two hours. This indicates that the technology increased the efficiency of the class significantly even though the students were engaged in the same type of activities.

*Student Surveys*

The final source of evidence for this study is the surveys students completed at the end of the semester. The surveys were reviewed by the authors of this study, and each responses were judged to be either favorable toward the inclusion of technology (F), unfavorable toward the inclusion of technology (U), or neutral toward the inclusion of technology (N). All responses were tabulated, the overall results can be seen in Table IV, and specific quotes which supported the general response patterns were collected.

**Table IV. Results from student survey on role of technology in physics class.**

|  | 2004 | 2005 |
|---|---|---|
| Favorable | 20.4% | 76.0% |
| Unfavorable | 27.9% | 5.0% |
| Neutral | 51.7% | 19.0% |

Results from the 2004 survey showed ambivalence toward technology, students equally wished there were more technology and thought the technology was adequate. This is not surprising, as these students had not experienced a technology-enhanced class, so were unable to evaluate the benefits of the technology. Further these students were pleased with their learning experience in the physics class and generally did not see how technology would have improved their experience. An example of this type of student response from 2004, "I don't know the answer to this. I can't imagine the class any other way. I don't know how equipment could benefit us, because we learned so adequately without the equipment. The class was fine without the extra equipment!"

*2005 Student Survey Results*

The survey results during Fall 2005 were far more favorable toward the inclusion of technology. As can be seen in Table 4, the 2005 students saw the benefits of technology in the class. This is not to say they did not have some reservations, which primarily stemmed from technical difficulties with the computers, including lack of wireless, computers that froze at inopportune times and difficulties with setting up the labs. But the overwhelming sentiment was that the computers improved the educational experience in important ways. One student described the role of the computers as essential in helping to understand experiments and that the computers played a role in enhancing group interactions, saying, "It was very effective in me understanding the experiments and results especially with the whole group looking at the results." Another student also pointed to the efficiency that the computers added saying, "I'm not sure that we could have gotten the idea if we hadn't had the computers. It would have been too much time wasted if we hadn't had them."

## Conclusions and discussion

This study exploited the clean experimental design of making only one systematic change to the instruction in introductory mechanics, and then collected three different forms of data to substantiate the changes in student learning when real-time data collection and analysis hardware and software was introduced. The data points to significant gains in leaning, and student responses to the inclusion of technology were very supportive. From this, we attribute the gains in learning to the inclusion of the technology. This finding is significant, because it measures increases in learning across an entire class as result from inclusion of technology.

However, the authors would not point to the technology as a panacea, instead the learning gains from the inclusion of technology are the result of technology being used in a way that supports a student-centered, active engagement curriculum. It would not be expected that a standard lecture-based mechanics course that switches from low-tech, traditional labs to high-tech traditional labs would have the same increase in learning gains. Instead these increases in learning exist because the student-centered pedagogy pre-dated the inclusion of the technology. Following are some ideas about the role of the technology in the pedagogy, which lead to improved learning.

The first and most obvious difference between the classes is the efficiency of using technology to collect and represent data. Students in the technology-enhanced class have data instantly represented for them, which allows them to focus immediately on the interpretation and analysis. Students lacking technology see a large time delay between conducting the investigation and beginning to interpret and analyze the data. Further, the lack of technology requires students to focus on creating the representations themselves, leaving them to question the accuracy of the representations. Students in the technology enhanced class were seen by the ethnographer discussing the meaning of the graphs while they were still involved in making predictions, which indicates that these students were engaging in analysis and prediction instantly rather than waiting. This time delay from phenomena to representation and analysis seems significant and worthy of further study.

The efficiency factor comes into play in other ways as well, because the labs are much faster, additional time is available. The additional time in this class was not dedicated to covering greater quantities of material, but instead to further investigation of the concepts and added problem solving. This can be seen as the reason for greater learning, and it should be. The technology can be used to help students achieve more by doing things that do not add to student learning more quickly, thereby freeing up more time for the activities that are important in learning.

The third factor in the role of technology in learning is how the technology is used to support active-engagement student-centered pedagogy. In this example, the students were deeply engrossed in the modeling curriculum, which focuses on having students build models and to validate the models with evidence. Because the students were able to use the technology to provide evidence, which reinforced the models they had developed, the experience was valuable to their understanding. The students had seen the technology as a valuable tool for developing modeling phenomena, and the models as the essential knowledge construct for understanding physics.

In short, this study has made a measurement of the impact that technology can have on student learning. The impact is significant but will be realized when the technology is used as a tool to support active-engagement in the understanding of physics.

## Acknowledgements

The authors would like to thank the National Science Foundation, this work is the direct result of NSF-DUE #0411344. Also Dr. Brewe would like to thank Hawaii Pacific University, especially Dr. Andrew Brittain, Dr. Martha Sykes and Dr. Valentina Abordonado.